\documentclass[aps,prl,preprint,groupedaddress]{revtex4}

\usepackage{epsfig}
\usepackage{amssymb}
\usepackage{times}

\begin{document}

\newcommand{\re}{\mathop{\mathrm{Re}}}
\newcommand{\im}{\mathop{\mathrm{Im}}}
\newcommand{\I}{\mathop{\mathrm{i}}}
\newcommand{\D}{\mathop{\mathrm{d}}}
\newcommand{\E}{\mathop{\mathrm{e}}}

\def\lambar{\lambda \hspace*{-5pt}{\rule [5pt]{4pt}{0.3pt}} \hspace*{1pt}}

{\Large  DESY 16-243}

{\Large  December 2016}

\bigskip

\bigskip

\bigskip

\bigskip

\bigskip


\title{
First operation of a harmonic lasing self-seeded free electron laser}



\author{E.A.~Schneidmiller}
\email[]{evgeny.schneidmiller@desy.de}
\author{B.~Faatz}
\author{M.~Kuhlmann}
\author{J.~R\"onsch-Schulenburg}
\author{S.~Schreiber}
\author{M.~Tischer}
\author{M.V.~Yurkov}
\affiliation{Deutsches Elektronen-Synchrotron (DESY),
Notkestrasse 85, D-22607 Hamburg, Germany}

\date{\today}

\begin{abstract}
Harmonic lasing is a perspective mode of operation of X-ray FEL user facilities that allows to
provide brilliant beams of higher energy photons for user experiments.
Another useful application of harmonic lasing is so called
Harmonic Lasing Self-Seeded Free Electron Laser (HLSS FEL)
that allows to improve spectral brightness of these facilities.
In the past, harmonic lasing
has been demonstrated in the FEL oscillators in infrared and visible wavelength ranges, but not in high-gain FELs and not
at short wavelengths. In this paper we report on the first evidence of the harmonic lasing and
the first operation of the HLSS FEL at the soft X-ray FEL user facility FLASH in
the wavelength range between 4.5 nm and 15 nm. Spectral brightness was improved in comparison with Self-Amplified
Spontaneous emission (SASE) FEL by a factor of six in the exponential gain regime. A better performance of HLSS FEL
with respect to SASE FEL in the post-saturation regime with a tapered undulator was observed as well.
The first demonstration of harmonic lasing in a high-gain FEL and at short wavelengths paves the way for a variety of
applications of this new operation mode in X-ray FELs.
\end{abstract}


\maketitle


\section{Introduction}

Successful operation of X-ray free electron lasers (FELs) \cite{flash,lcls,sacla}, based on
self-amplified spontaneous emission (SASE) principle \cite{ks-sase},
down to an $\rm{\AA}$ngstr{\"o}m regime opens up new horizons for photon science. Even shorter
wavelengths are requested by the scientific community.

One of the most promising ways to extend the photon energy range of high-gain X-ray FELs is to use harmonic
lasing which is
the FEL instability at an odd harmonic of the
planar undulator \cite{murphy,hg-2,kim-1,mcneil,sy-harm} developing independently from the lasing at the fundamental.
Contrary to the nonlinear harmonic
generation \cite{hg-2,hg-3,kim-1,harm-prst,hg-exp-1,flash,lcls-harm}
(which is driven by the fundamental in the vicinity of saturation),
harmonic lasing can provide much more intense, stable, and narrow-band radiation
if the fundamental is suppressed. The most attractive feature of saturated harmonic lasing
is that the spectral brightness of a harmonic is comparable to that of the fundamental \cite{sy-harm}.

Another interesting option, proposed in \cite{sy-harm}, is the possibility to improve spectral brightness of an X-ray FEL by
the combined lasing on a harmonic in the first part of the undulator (with an increased undulator parameter K)
and on the fundamental in the second part of the undulator. Later this concept was named Harmonic Lasing Self-Seeded
FEL (HLSS FEL) \cite{hlss}. Even though this scheme is not expected to provide an ultimate monochromatization of the FEL radiation
as do self-seeding schemes using optical elements \cite{ss-soft,ss-mev,ss-wake},
it has other advantages that we briefly discuss below in the paper.

Harmonic lasing was initially proposed for FEL oscillators \cite{colson} and was tested experimentally in infrared
and visible wavelength
ranges \cite{benson-madey,warren,hajima,sei}. It was, however, never demonstrated in high-gain FELs and at a short wavelength.
In this paper we present the
first successful demonstration of this effect at the second branch of the soft X-ray FEL user facility FLASH \cite{fl2}
where we managed to run HLSS FEL in the wavelength range between 4.5 nm and 15 nm.

\section{Harmonic lasing}

Harmonic lasing in single-pass high-gain FELs \cite{murphy,hg-2,kim-1,mcneil,sy-harm} is the
amplification process of higher odd harmonics developing independently of each
other (and of the fundamental harmonic) in the exponential gain regime. In
the case of a SASE FEL the fluctuations of the beam current with frequency components
in the vicinity of a wavelength

\begin{equation}
\lambda_h = \frac{\lambda_{\mathrm{w}} (1+K^2)}{2h \gamma^2}    \ \ \ \ \ \ \ \ \   h = 1,3,5...
\label{lambda}
\end{equation}

\noindent serve as an input signal for amplification process.
Here $\lambda_{\mathrm{w}}$ is the undulator period, $\gamma$ is relativistic
factor, $h$ is harmonic number, and $K$ is the rms undulator parameter:

\begin{displaymath}
K = 0.934 \ \lambda_{\mathrm{w}} [{\mathrm{cm}}] \ B_{\mathrm{rms}} [{\mathrm{T}}] \ ,
\end{displaymath}

\noindent $B_{\mathrm{rms}}$ being the rms undulator field (peak field divided by $\sqrt{2}$ for a planar undulator
with the sinusoidal field).

An advantage of harmonic lasing over lasing on the fundamental at the same wavelength can be demonstrated
for the case of a gap-tunable undulator. In this case one uses a higher K-value for harmonic lasing, i.e. for the
lasing on the fundamental one has to reduce K to the value $K_{re}$:

\begin{equation}
K_{re}^2 = \frac{1+K^2}{h} -1  \ .
\label{ret-k-text}
\end{equation}

\noindent Obviously, $K$ must be larger than $\sqrt{h-1}$.

Then one can derive a ratio of the gain length of the fundamental, $L_{g}^{(1)}$, to the gain length of a harmonic
$L_{g}^{(h)}$ \cite{sy-harm}:

\begin{equation}
\frac{L_{g}^{(1)}}{L_{g}^{(h)}} =   \frac{h^{1/2} K A_{JJh}(K)}{K_{re} A_{JJ1}(K_{re})}   \ .
\label{ratio-1h-3d-text}
\end{equation}

\begin{figure*}[tb]

\includegraphics[width=.6\textwidth]{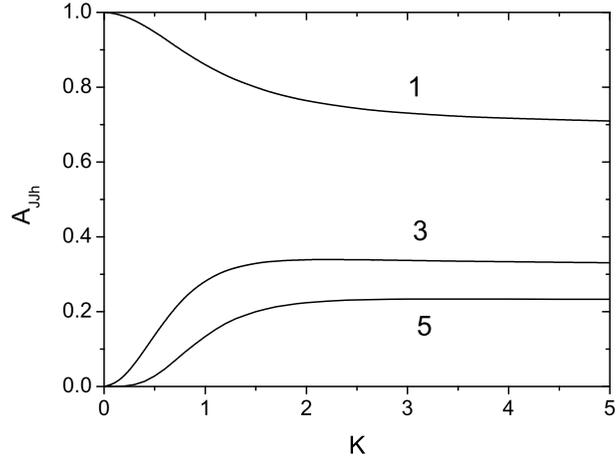}

\caption{\small Coupling factors for the 1st, 3rd, and 5th harmonics (denoted with 1, 3, and 5, correspondingly) versus
rms undulator parameter.}

\label{ajj}
\end{figure*}

\noindent Here
$A_{JJh}(K) = J_{(h-1)/2} \left( \frac{hK^2}{2(1+K^2)} \right) - J_{(h+1)/2} \left( \frac{hK^2}{2(1+K^2)} \right)$
is the coupling factor for harmonics with $J_n$ being Bessel functions. The coupling factors for the 1st, 3rd, and 5th
harmonics are shown in Fig.~\ref{ajj}.

The formula (\ref{ratio-1h-3d-text}) is obtained in the frame of the three-dimensional theory including diffraction of
the radiation,
emittance, betatron motion (and for an optimized beta-function) but assuming a negligible energy spread. The plot
of the ratio of gain lengths (\ref{ratio-1h-3d-text}) is presented in Fig.~2. It is clearly seen that harmonic lasing
has always a shorter gain length under above mentioned conditions (and the ratio is larger than that obtained in
one-dimensional model \cite{mcneil}).
The ratio shown in Fig.~2 starts to diverge rapidly for the values of K
approaching $\sqrt{2}$, and lasing at the fundamental becomes impossible
below this point. However, there still remains a reserve in the value of
parameter K allowing effective lasing at the third harmonic.

Amplification process of harmonics degrades with the increase of the energy
spread in the electron beam more rapidly than that of the fundamental. However,
in practical situations there is always the range of parameters for which the harmonic
lasing still has an advantage \cite{sy-harm}.

\begin{figure*}[tb]

\includegraphics[width=.6\textwidth]{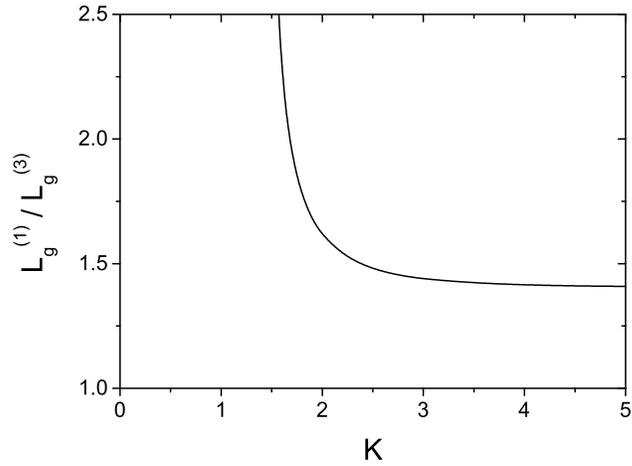}

\caption{\small Ratio of the gain length of the retuned fundamental to the gain length of the third harmonic
(\ref{ratio-1h-3d-text})
for lasing at the same wavelength versus rms undulator parameter $K$. The ratio is derived in the frame of
the three-dimensional theory
for an optimized beta-function and negligible energy spread \cite{sy-harm}.}

\label{gl-3d}
\end{figure*}

The most attractive feature of the saturated harmonic lasing
is that the spectral brightness (or brilliance)
of harmonics is comparable to that of the fundamental \cite{sy-harm}.
Indeed, a good estimate for the
saturation efficiency is $\lambda_{\mathrm{w}}/(h L_{\mathrm{sat},h})$, where
$L_{\mathrm{sat},h}$ is the saturation length of a harmonic ($h=1$ for the fundamental).
At the same time, the relative rms bandwidth
has the same scaling. In other words, reduction of power is compensated by the bandwidth reduction, and the spectral
power remains the same.
If we consider the lasing at the same wavelength on the fundamental and on a harmonic (with the retuned undulator
parameter $K$), transverse coherence properties are about the same since they are mainly
defined by the emittance-to-wavelength ratio \cite{coherence-oc, coherence-njp}.
Thus, also the spectral brightness is about the same in both cases.

Although known theoretically for a long time \cite{murphy,hg-2,kim-1,mcneil}, harmonic lasing in high-gain FELs was never
demonstrated experimentally. Moreover, it was never considered for practical applications in X-ray FELs.
The situation was changed after publication of ref.~\cite{sy-harm} where it was concluded that
the harmonic lasing in X-ray FELs is much more robust than usually thought, and can be effectively used in
the existing and future X-ray FELs. In particular, the European XFEL \cite{euro-xfel-tdr}
can greatly outperform
the specifications in terms of the highest possible photon energy: it can reach 60-100 keV range for the
third harmonic lasing.
It was also shown \cite{cw} that one can keep sub-$\rm{\AA}$ngstr{\"o}m range of operation of the European XFEL
after CW upgrade
of the accelerator with a reduction of electron energy from 17.5 GeV to 7 GeV.
Another application of harmonic lasing is a possible upgrade of FLASH \cite{flash-1-2} with the aim to
increase the photon energy
up to 1 keV with the present energy 1.25 GeV of the accelerator. To achieve this goal, one should install a specially
designed undulator optimized for the third harmonic lasing as suggested in \cite{fl-harm-las}.

\section{Harmonic lasing self-seeded FEL}

A poor longitudinal coherence of SASE FELs \cite{bon-rho,stat,book} stimulated efforts for its improvement.
Since an external seeding seems to be difficult to realize in X-ray regime, a so called self-seeding has been proposed
\cite{ss-soft,ss-mev,ss-wake}.
There are alternative approaches for reducing bandwidth and increasing spectral brightness of X-ray FELs without using
optical elements. One of them \cite{isase,hbsase}
suggests to use chicanes inside the undulator system to increase slippage of
the radiation and to establish long-range correlations in the radiation pulse.
Another method was proposed in \cite{sy-harm} and is based on the combined lasing on a harmonic in
the first part of the undulator
(with increased undulator parameter K, see formula (\ref{ret-k-text}))
and on the fundamental in the second part. In this way the second part of the undulator is seeded by a narrow-band signal
generated via a harmonic lasing in the first part.
This concept was named HLSS FEL (Harmonic Lasing Self-Seeded FEL) \cite{hlss}.
Note that a very similar concept was proposed in \cite{psase} and was called a purified SASE FEL, or pSASE.

\begin{figure}[tb]

\includegraphics[width=.8\textwidth]{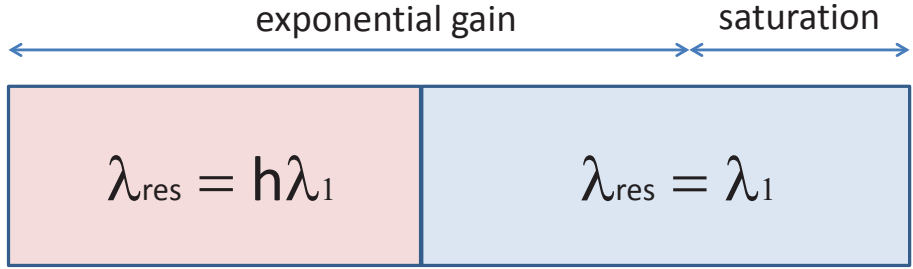}

\caption{\small Conceptual scheme of a harmonic lasing self-seeded FEL}

\label{hlss}
\end{figure}

Typically, gap-tunable undulators are planned to be used in  X-ray FEL facilities. If maximal undulator parameter $K$ is
sufficiently large, the concept of harmonic lasing self-seeded FEL can be applied in such undulators (see Fig.~\ref{hlss}).
An undulator is divided into two
parts by setting two different undulator parameters such that the first part is tuned to a $h$-th sub-harmonic
of the second part which is tuned to a wavelength of interest $\lambda_1$.
Harmonic lasing occurs in the exponential gain regime in the first part of the
undulator, also the fundamental in the first part stays well below
saturation. In the second part of the undulator the fundamental is resonant to the
wavelength, previously amplified as the harmonic. The amplification
process proceeds in the fundamental up to saturation. In this case the
bandwidth is defined by the harmonic lasing (i.e. it is reduced by a
significant factor depending on harmonic number) but the saturation power is
still as high as in the reference case of lasing at the fundamental in the whole undulator, i.e.
the spectral brightness increases.

The enhancement factor of the coherence length (or, bandwidth reduction factor), that one obtains in HLSS FEL in comparison with a
reference case of lasing in SASE FEL mode in the whole undulator, reads \cite{hlss}:

\begin{equation}
R \simeq h \ \frac{\sqrt{L_{\mathrm{w}}^{(1)} L_{\mathrm{sat},h}}}{L_{\mathrm{sat},1}}
\label{R}
\end{equation}

\noindent Here $h$ is harmonic number, $L_{\mathrm{sat},1}$ is the saturation length in the reference case of
the fundamental lasing with the lower K-value, $L_{\mathrm{w}}^{(1)}$ is the length of the first part of the undulator,
and $L_{\mathrm{sat},h}$ is the saturation length of harmonic lasing.
We notice that it is beneficial to increase the length of the first part of the undulator. Since it must be
shorter than the saturation length of the fundamental harmonic in the first section, one can consider delaying the saturation
of the fundamental with the help of phase shifters \cite{mcneil,sy-harm} in order to increase $L_{\mathrm{w}}^{(1)}$.
However, for the sake of simplicity, we did not use this option in our experiments.

Despite the bandwidth reduction factor (\ref{R}) is significantly smaller than that of self-seeding schemes using
optical elements \cite{ss-soft,ss-mev,ss-wake}, the HLSS FEL scheme is very simple and robust, and it does not require any additional installations,
i.e. it can always be used in existing or
planned gap-tunable undulators with a sufficiently large K-value.

One more advantage of the HLSS FEL scheme over the SASE FEL (and in many cases over a self-seeded FEL) is the possibility of
a more efficient use of a post-saturation taper \cite{kmr,fawley-1,arctan}
for an improved conversion of the electron beam power to the
FEL radiation power \cite{hlss,hlss-ipac16}. It is well-known \cite{fawley-2} that a seeded (self-seeded) FEL works better in the post-saturation tapering regime than
SASE FEL. In the latter case, a poor longitudinal coherence limits the length of the tapered part of the undulator to a length on
which a slippage of the radiation with respect to the electron beam is comparable with the FEL coherence length
\cite{taper-spie,taper-fel15}. In a self-seeded FEL the
coherence length is much larger and it does not limit the performance of the tapered FEL (unless a sideband instability starts
playing a role \cite{kmr}). A disadvantage of a self-seeded FEL is that the saturation length is almost doubled
with respect to the
SASE regime, so that the available length for tapering the undulator may become too short. Considering now the HLSS FEL, we can state
that it combines both advantages: coherence length is significantly larger than in the case of the SASE FEL, and the saturation length
is shorter than that of the SASE FEL. In other words, there is more undulator length, available
for tapering, than in the cases of the self-seeded FEL and SASE, and the longitudinal coherence is good enough to perform efficient
tapering over this length. This makes us believe that HLSS FEL will become a standard mode of operation of X-ray FEL
facilities.

Numerical simulations of the HLSS FEL were presented in \cite{hlss} for the European XFEL \cite{euro-xfel-tdr}
and in \cite{hlss-ipac16} for FLASH \cite{flash-1-2}.
In this paper we report on the first operation of the harmonic lasing self-seeded FEL.
The experiment was performed at the 2nd undulator line of the free electron laser FLASH \cite{flash,fl2,flash-1-2}.
We detected clear evidence of the 3rd harmonic lasing in the
wavelength range from 4.5 nm to 15 nm and compared performance of HLSS FEL and SASE FEL.
Obtained experimental results are in good agreement with
expectations \cite{hlss,hlss-ipac16}: HLSS FEL provides more powerful photon beams
with improved longitudinal coherence.

\section{Operation of the HLSS FEL at FLASH2}

The first soft X-ray FEL user facility
FLASH \cite{flash,flash-1-2} was upgraded to split the electron pulse trains between the two
undulator lines so that
the accelerator with maximum energy of 1.25 GeV now drives both lines.
In a new separate tunnel, a second undulator line, called FLASH2, with a variable-gap undulator was installed,
while a new experimental hall has space for up to six experimental stations \cite{fl2}.
The gap-tunable undulator of FLASH2 consists of twelve 2.5 m long sections with the undulator period of 3.14 cm and
the maximum rms K-value about 1.9. This makes it possible (see formula (\ref{ret-k-text})) to study the HLSS FEL
scheme with the 3rd harmonic seeding.
Due to the parallel
operation with FLASH1 undulator line, the electron beam diagnostics, placed in the common part of the machine,
was not available during the measurements.
Moreover, FLASH2 is not equipped with the longitudinal beam diagnostics yet.
For this reasons we can not directly compare our measurements with numerical simulations.
We could, however, observe a good qualitative agreement with the simulations \cite{hlss-ipac16}
that were done before the measurements.

\subsection{First lasing at 7 nm}

On May 1, 2016 we were able to successfully perform the first test of HLSS FEL at FLASH2.
Electron energy was 948 MeV,
charge 0.4 nC. Initially we tuned 10 undulator sections to a standard SASE,
operating in the exponential gain regime at the wavelength of 7 nm
(rms K parameter was 0.73);
the pulse energy was 12 $\mu$J. Then we detuned the first section,
tuned it to the third subharmonic (rms K was 1.9) and scanned it around 21 nm.
We repeated the measurements with the first two sections,
and then with the first three sections. Note that the fundamental at 21 nm was also in the exponential gain regime,
pulse energy after three undulator sections was 40 nJ, i.e. it was far away from saturation (which was achieved at
the 200 $\mu$J level).
This means, in particular, that the
nonlinear harmonic generation in the first part of the undulator is excluded.

\begin{figure*}[tb]

\includegraphics[width=.6\textwidth]{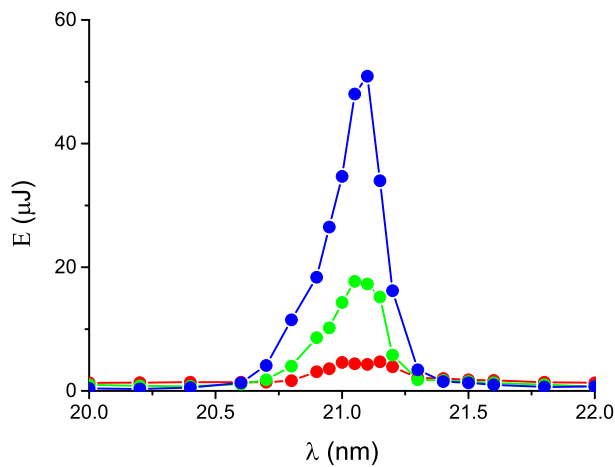}

\caption{\small Scan of the resonance wavelength of the first part of the undulator consisting of one undulator section (red),
two sections
(green), and three sections (blue). Pulse energy is measured after the second part of the undulator tuned to 7 nm.
}

\label{3d-gam-k}
\end{figure*}

One can see from Fig.~\ref{3d-gam-k} that the effect is essentially resonant. For example,
in the case when three undulator sections
were scanned, the ratio of pulse energies at the optimal tune, 21.1 nm, and at the tune of 20 nm is 51 $\mu$J/0.3 $\mu$J = 170.
This ratio is likely underestimated because the background radiation at the fundamental at 20 nm (even being much
weaker, about 40 nJ) is more efficiently
detected by the microchannel plate (MCP) based detector \cite{mcp,mcp-2} used in this measurement. Note that the MCP detector has a
very large dynamical range and a high signal-to-noise ratio. For these reasons it is best suited to measurements of
the FEL gain curve and statistical properties of the FEL radiation \cite{flash,behrens,coh-ipac16}.
This detector has no absolute
calibration, therefore
in our experiments we used gas monitor detector (GMD) \cite{gmd,njp-flash} to absolutely calibrate the MCP detector at the level of
10 $\mu$J.

We claim that there can be only one explanation of the effect that we observe in
Fig.~\ref{3d-gam-k}:
FEL gain at 7 nm is strongly reduced as soon as the first part of the undulator is detuned, and then the gain is recovered (and
becomes even larger) due to the 3rd harmonic lasing in the first part as soon as the resonant wavelength is 21 nm.

We should stress that the pulse energy with three retuned undulator sections (51 $\mu$J) is significantly larger than that in
the homogeneous undulator tuned to 7 nm (it was 12 $\mu$J). This is because the gain length of harmonic lasing
is shorter than
that of the fundamental tuned to the same wavelength
(see formula (\ref{ratio-1h-3d-text}), Fig.~2 and refs. \cite{mcneil,sy-harm,hlss,hlss-ipac16}).

\subsection{Improvement of the longitudinal coherence at 11 nm}

We continued the studies of the HLSS FEL at FLASH2 in June 2016. Since the electron energy was different (757 MeV),
we lased at another wavelength, 11 nm. We also used a different charge, 0.25 nC, in this experiment.
The undulator settings were similar to the previous case: we used ten undulator
modules, rms K-parameter was 0.73 in SASE mode and 1.9 in the first part of the undulator in HLSS mode. The difference with
the previous measurements was that we detuned four undulator modules in HLSS regime.

In the same way as in the previous experiment, we performed the scan of the K parameter in the first part of the undulator and
saw a resonance behavior again. In combination with the fact that the fundamental at 33 nm was by three orders
of magnitude below saturation this proves that we had harmonic lasing at 11 nm in the first part of the undulator.
The pulse energies were 11 $\mu$J in SASE mode and 53 $\mu$J in HLSS mode.

The main goal of this run was to demonstrate that HLSS scheme indeed helps to improve the longitudinal coherence of FEL pulses with
respect to the standard SASE regime. One can do this by the demonstration of the bandwidth reduction and by the measurements
of an increase of the coherence time.

\begin{figure}[tb]

\includegraphics[width=.6\textwidth]{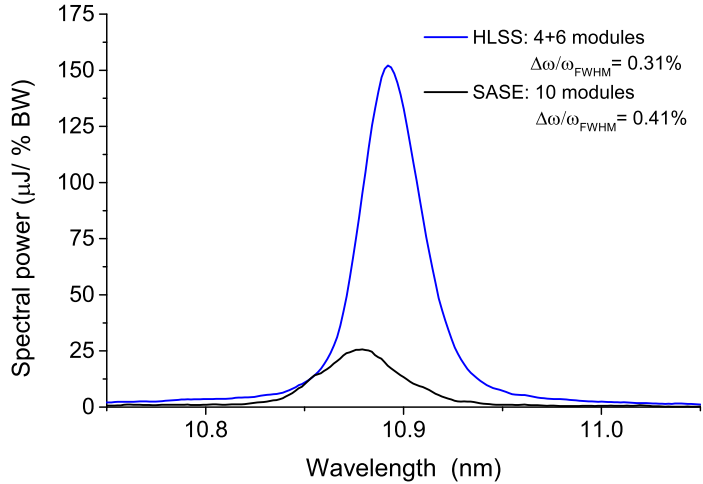}

\caption{\small
Spectral density of the radiation energy for HLSS FEL configuration (blue) and for SASE FEL (black).
}
\label{spectr-11nm}
\end{figure}

The spectra were measured with the wide-spectral-range XUV spectrometer \cite{spectr} of
FLASH2.  A narrow entrance slit is imaged by a 1200 l/mm spherical
variable line spacing grating in the 5th grating order which allows for a
resolution better than 0.01 nm.
In Fig. \ref{spectr-11nm} we
present the averaged spectra for two study cases: SASE FEL with ten undulator modules and HLSS FEL with four modules
tuned to 33 nm and six modules tuned to 11 nm. Let us note that a per cent level discrepancy between the measured
wavelength (about 10.9 nm) and the wavelength expected by the undulator server (11 nm) comes from the fact that the server
uses electron energy calculated from the RF vector sum and not from a direct measurement of the electron beam energy.
However, the red shift of the radiation for the HLSS configuration with respect to the SASE case is real and can be
explained by the fact that a seeded FEL radiates more efficiently in the case of a small red shift \cite{book}.

The spectra in Fig.~\ref{spectr-11nm} are the results of averaging over 50 single-shot spectra in each case.
One can see that HLSS FEL indeed has a smaller bandwidth, 0.31\%, as compared to 0.41\% in the case of SASE FEL. The bandwidth
reduction factor is 1.3 from this measurement. The spectral power, however, differs by a factor of six due to an additional
increase of pulse energy in HLSS regime. This happens because the 3rd harmonic lasing at 11 nm has a shorter gain length
than lasing at the same wavelength on the fundamental.

An expected bandwidth reduction factor (or coherence enhancement factor) $R$
from formula (\ref{R}) can be estimated at 1.7.
The discrepancy can in a general case be explained by the energy jitter and/or energy chirp in the electron beam.
The energy jitter effect is supposed to give a small contribution to the spectrum broadening since the FLASH accelerator
was quite stable during the measurement, the energy stability can be estimated at the level of a few $10^{-4}$.
A contribution of the energy
chirp, however, being converted to a frequency chirp within an FEL pulse, can be significant. The energy chirp appears in the
accelerator on the one hand due to off-crest acceleration, needed for the bunch compression in magnetic chicanes, and on the
other hand due to collective self-fields in the bunch (wakefields, longitudinal space charge) \cite{flash}.
Both contributions can partially
or fully compensate each other, this depends on accelerator settings. In the experiment we could
tweak the bunch compression, trying to minimize the HLSS FEL bandwidth, and we succeeded partially.

\begin{figure}[tb]

\includegraphics[width=.45\textwidth]{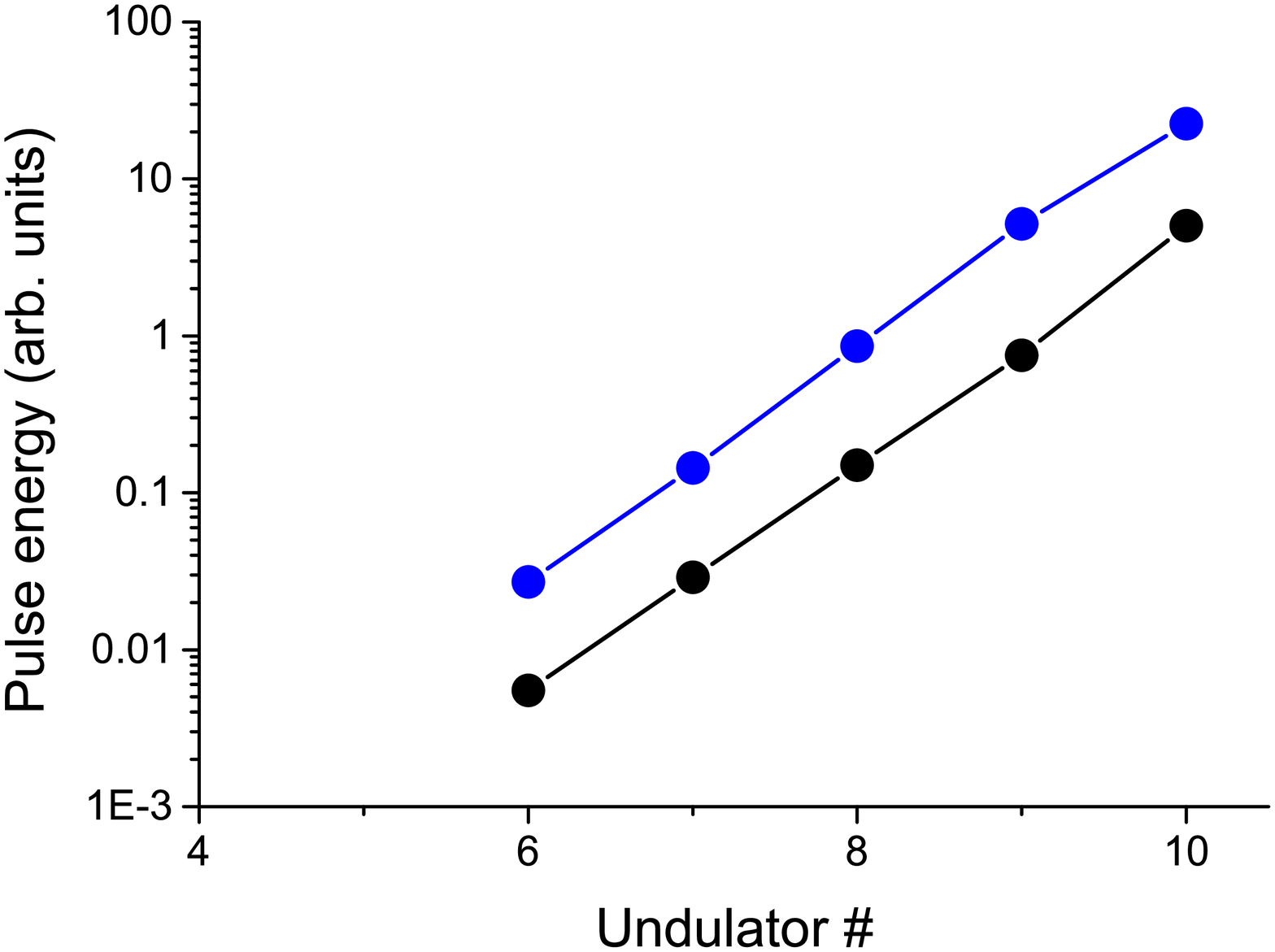}
\includegraphics[width=.45\textwidth]{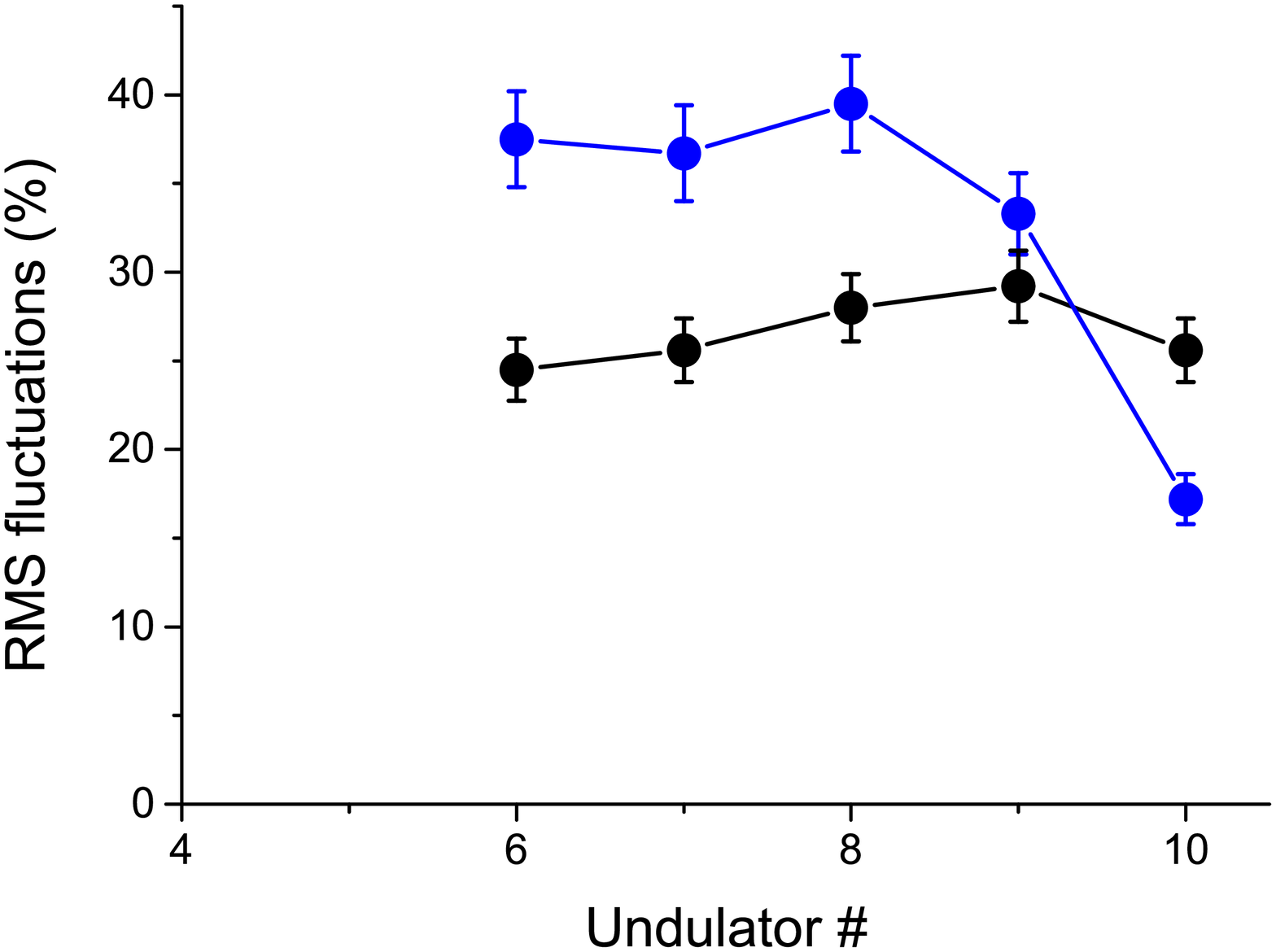}

\caption{\small
Radiation pulse energy (left plot) and pulse energy fluctuations (right plot)
in the second part of the undulator
for HLSS (blue) and for SASE (black).
Small aperture in front of the MCP detector is used in this measurement.
}
\label{gain-curve}
\end{figure}

Another method of determination of an improvement of the longitudinal coherence
(independent of the presence of the frequency chirp in FEL pulses) is based on statistical measurements
of the FEL pulse energy along the undulator length. It is well known \cite{stat,book} that in
high-gain linear regime the radiation from a SASE FEL has a statistics of a completely
chaotic polarized light \cite{goodman}.
Shot-to-shot rms fluctuations of the FEL pulse energy $\sigma$ are connected with the number of modes by a
simple relation: $M = 1/\sigma^2$.
Number of modes can be represented as a product of the numbers of longitudinal, $M_L$, and transverse, $M_T$, modes. The latter
is usually close to one, $M_T \simeq 1.1 - 1.2$ when a SASE FEL is well designed and optimized
\cite{coherence-oc,coherence-njp}.
If one uses a small aperture to select only the central part of the FEL beam, the pulse energy fluctuations are a measure of
the number of the longitudinal modes \cite{coh-ipac16} : $M_L = 1/\sigma^2$.
For a given FEL
pulse length, the coherence length $L_{coh}$ is inversely proportional to the number of the longitudinal modes, $M_L$. Making
a reasonable assumption that the FEL pulse length is the same in both cases, HLSS and SASE, we
end up with a simple ratio of coherence lengths for these two cases:

\begin{equation}
R = \frac{L_{coh}^{\small HLSS}}{L_{coh}^{\small SASE}} \simeq  \frac{M_L^{\small SASE}}{M_L^{\small HLSS}} =
\frac{\sigma_{\small HLSS}^2}{\sigma_{\small SASE}^2}
\label{coh-length}
\end{equation}

In Fig.~\ref{gain-curve} we present the measurements of the FEL pulse energy and its fluctuations versus undulator length
for a small aperture
(significantly smaller than the FEL beam size) in front of the MCP detector. The measurements start behind the sixth undulator
section because at this position the contribution of the background radiation at 33 nm is already negligible. In both cases,
HLSS and SASE, the maximum of pulse energy fluctuations is achieved within the part of the undulator where the measurements
were performed. However, in HLSS case the fluctuations drop down more significantly because the FEL enters nonlinear
stage of amplification in this case. As one can see, in the linear regime of the FEL operation (sections 6 to 8) the
fluctuations for HLSS case are visibly larger than in the SASE case. The validity of an assumption that pulse length
in both cases is the same is justified by the fact that both FEL configurations were driven by the same electron beam under
the same conditions. We did the measurements twice for each configuration to make sure
that the results are not affected by any drifts in the accelerator. Thus, we can conclude that in the HLSS case we had a
smaller number of modes, or a larger coherence length. Using formula (\ref{coh-length}) with the fluctuations measured behind the
8th undulator section for HLSS and the 9th section for SASE,
we obtain an estimate for the coherence enhancement factor in the end of the exponential gain regime:
$R \simeq 1.8 \pm 0.3$. This is in a
good agreement with already presented theoretical estimate $R \simeq 1.7$ obtained from (\ref{R}).

Note that this moderate enhancement, observed in our experiment, is obtained because we are limited to application
of the third (and not higher) harmonic at FLASH2.
Further improvement can be done by increasing the length of the first part of the undulator (see formula (\ref{R})),
making sure that
the fundamental in the first part stays well below saturation
(one can delay the saturation by using phase shifters as suggested in \cite{mcneil,sy-harm}).
In a gap-tunable undulator with a higher $K$, like SASE3 undulator of the European XFEL (with the rms $K$ about 7),
one can, in principle, use a much higher harmonic number thus expecting a much higher coherence enhancement factor.

\subsection{A more efficient post-saturation taper at 15 nm}

In November 2016 we set up HLSS FEL as a configuration with four first undulators tuned to 45 nm and the
last eight undulators tuned to 15 nm. The electron energy was 645 MeV, the charge was 100 pC, the rms value of K was 1.9 in
the first part of the undulator and 0.73 in the second part. We reached FEL saturation in
SASE and HLSS modes, and applied post-saturation taper to improve FEL efficiency \cite{kmr,fawley-1,arctan}.

\begin{figure}[tb]

\includegraphics[width=.6\textwidth]{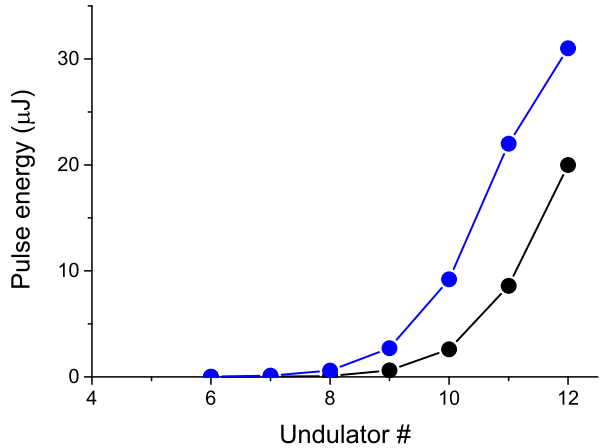}

\caption{\small
Radiation pulse energy versus position
in the undulator
for HLSS (blue) and for SASE (black).
Post-saturation taper was optimized for both cases.
}
\label{hlss-taper}
\end{figure}

Post-saturation taper in FLASH2 undulator is implemented as a step-taper (i.e. the undulator K changes from
section to section but is constant within a section) with linear or quadratic law. We used
quadratic taper and for each mode (HLSS and SASE) optimized two parameters: beginning of tapering and the taper
depth. We ended up with the following optimized parameters: beginning of tapering was in the 9th (10th) undulator
and the taper depth was 0.9\% (0.7\%) for HLSS (SASE). Pulse energy was enhanced for HLSS configuration
from 18 $\mu$J in non-tapered undulator to 31 $\mu$J when post-saturation taper was applied. In case of SASE FEL the
respective enhancement was from 15 $\mu$J to 20 $\mu$J. The pulse energy versus undulator length for both
operation modes is presented in Fig.~\ref{hlss-taper}.

Note that a similar efficiency enhancement was previously observed in numerical simulations \cite{hlss,hlss-ipac16}.
As it was discussed above, the improvement of post-saturation taper regime is achieved in HLSS case for two reasons:
an earlier saturation and a better longitudinal coherence than in SASE case.

\begin{figure*}[tb]

\includegraphics[width=.6\textwidth]{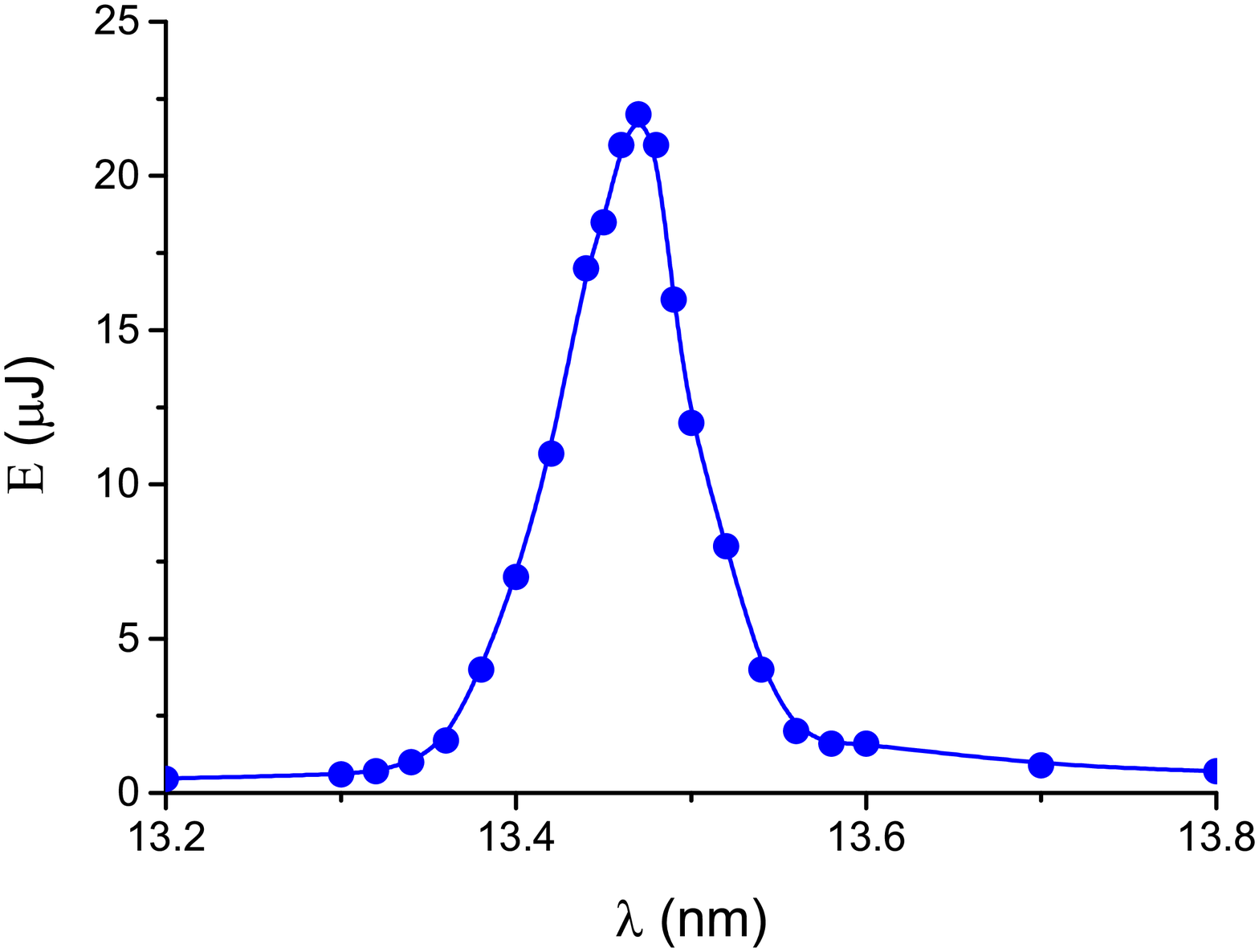 }

\caption{\small Scan of the resonance wavelength of the first part of the undulator consisting of three undulator sections.
Pulse energy is measured after the second part of the undulator tuned to 4.5 nm and operated close to the FEL saturation.
}

\label{k-scan-13dot5nm}
\end{figure*}

\subsection{Lasing at 4.5 nm}

In September 2016 we were able to drive HLSS FEL by the electron beam with a higher energy, 1080 MeV, and thus obtain lasing
at 4.5 nm in HLSS configuration. Initially, we tuned SASE regime with 12 active undulator sections (rms K value was 0.53),
and could establish an onset of saturation with pulse energy at the level of 20 $\mu$J.
Then we tuned first three sections to 13.5 nm (increasing rms K value to 1.69), thus providing the third harmonic signal
at 4.5 nm for seeding the last nine undulators. The scan of the undulator tune of the first three modules is presented in
Fig.~\ref{k-scan-13dot5nm}. The resonant behavior together with the fact that the fundamental at 13.5 nm was more than three orders of
magnitude below saturation proves that we had the third harmonic lasing at 4.5 nm in the first part of the undulator.

\section{Conclusion}

We were able to successfully demonstrate the harmonic lasing phenomena and the
HLSS FEL principle at FLASH2 in the wavelength range between 4.5 and 15 nm.
A change from SASE to HLSS configuration was very
simple and fast, it worked well independently of a wavelength and accelerator settings. We can, therefore, forecast that HLSS
may become a standard mode of operation of the X-ray FEL user facilities with gap-tunable undulators, providing an improvement
of the longitudinal coherence, a reduction of the saturation length and a possibility of a more efficient post-saturation
tapering.

It is also important to note that the first evidence of harmonic lasing in a
high-gain FEL and at a short wavelength (down to 4.5 nm) paves the way for a variety of applications of this effect in X-ray
FEL facilities \cite{sy-harm,hlss,fl-harm-las,cw}.

\section{Acknowledgments}

We are grateful to FLASH team for technical support. We would like to thank
R.~Brinkmann, J.~Schneider, E. Weckert and W. Wurth for valuable discussions and support of this work.

\end{document}